\documentclass[RNAAS]{aastex62}
\usepackage{color}

\usepackage{hyperref}
\hypersetup{colorlinks=true,linkcolor=blue,filecolor=magenta,urlcolor=blue}

\newcommand{\Gaia}{{\sl Gaia}}

\newcommand{\Lsun}{\mbox{$L_{\sun}$}}

\newcommand{\Mjup}{\mbox{$M_{\rm Jup}$}}

\newcommand{\degree}{\mbox{$^{\circ}$}}

\newcommand{\masyr}{\hbox{mas\,yr$^{-1}$}}

\newcommand{\Lbol}{\mbox{$L_{\rm bol}$}}

\newcommand{\Teff}{\mbox{$T_{\rm eff}$}}
\newcommand{\logg}{\mbox{$\log(g)$}}

\newcommand{\vlg}{\hbox{\textsc{vl-g}}}
\newcommand{\intg}{\hbox{\textsc{int-g}}}

\newcommand{\gps}{\ensuremath{g_{\rm P1}}}
\newcommand{\rps}{\ensuremath{r_{\rm P1}}}
\newcommand{\ips}{\ensuremath{i_{\rm P1}}}
\newcommand{\zps}{\ensuremath{z_{\rm P1}}}
\newcommand{\yps}{\ensuremath{y_{\rm P1}}}

\newcommand{\grizy}{\gps\rps\ips\zps\yps}

\newcommand{\etal}{et al.}

\newcommand{\obj}{VHS~J1256$-$1257}

\begin{document}

\title{The Parallax of VHS~J1256$-$1257 from CFHT and Pan-STARRS\,1}

\correspondingauthor{Trent J.\ Dupuy}
\email{tdupuy@gemini.edu}
\author[0000-0001-9823-1445]{Trent J.\ Dupuy}
\affiliation{Gemini Observatory, Northern Operations Center, 670 N.\ A'ohoku Place, Hilo, HI 96720, USA}
\author[0000-0003-2232-7664]{Michael C.\ Liu}
\affiliation{Institute for Astronomy, University of Hawaii, 2680 Woodlawn Drive, Honolulu, HI 96822, USA}
\author[0000-0002-7965-2815]{Eugene A.\ Magnier}
\affiliation{Institute for Astronomy, University of Hawaii, 2680 Woodlawn Drive, Honolulu, HI 96822, USA}
\author[0000-0003-0562-1511]{William M.\ J.\ Best}
\affiliation{The University of Texas at Austin, Department of Astronomy, 2515 Speedway C1400, Austin, TX 78712, USA}
\author[0000-0001-8365-5982]{Isabelle Baraffe}
\affiliation{University of Exeter, Physics and Astronomy, EX4 4QL Exeter, UK}
\affiliation{\`Ecole Normale Sup\'erieure, Lyon, CRAL (UMR CNRS 5574), Universit\'e de Lyon, France}
\author[0000-0002-8342-9149]{Gilles Chabrier}
\affiliation{\`Ecole Normale Sup\'erieure, Lyon, CRAL (UMR CNRS 5574), Universit\'e de Lyon, France}
\affiliation{University of Exeter, Physics and Astronomy, EX4 4QL Exeter, UK}
\author[0000-0003-0536-4607]{Thierry Forveille}
\affiliation{Universit\'e Grenoble Alpes, CNRS, IPAG, F-38000 Grenoble, France}
\author[0000-0003-3050-8203]{Stanimir A. Metchev}
\affiliation{The University of Western Ontario, Department of Physics and Astronomy, 1151 Richmond Avenue, London, ON N6A 3K7, Canada}
\author[0000-0001-6172-3403]{Pascal Tremblin}
\affiliation{Maison de la Simulation, CEA, CNRS, Univ. Paris-Sud, UVSQ, Universit\'e Paris-Saclay, F-91191 Gif-sur-Yvette, France}


\section{} 

VHS~J125601.92$-$125723.9 (\obj) comprises a nearly equal-flux  $0\farcs12$ binary (``AB'') and $8\farcs1$ companion (``b''), with spectral types of M$7.5\pm0.5$~\intg\ (combined-light) and L$7.0\pm1.5$~\vlg, respectively \citep{2015ApJ...804...96G,2016ApJ...818L..12S,2016ApJ...830..114R}. The \citet{2015ApJ...804...96G} parallactic distance of $12.7\pm1.0$\,pc indicates the companion is unusually faint relative to known young objects and may be planetary mass (11$^{+10}_{-2}$~\Mjup), but \citet{2016ApJ...818L..12S} infer a spectrophotometric distance of $17.2\pm2.6$\,pc for the binary host and a companion mass of up to 35\,\Mjup.  \Gaia~DR2 reports combined-light photometry for \obj{AB} but no parallax or proper motion and does not detect the companion.
 
We monitored the companion \obj{b} with the Canada-France-Hawaii Telescope (CFHT)
from 2016--2019. 
Using 20-s exposures in the $J$ band, we achieved $S/N=30-50$ on the target in individual frames, from which we measured the $(x,y)$ positions of it and 244 reference stars. (The primary \obj{AB} was saturated in our data, but it did not impact our measurements.) Using our custom pipeline \citep{2012ApJS..201...19D,2015ApJ...805...56D}, we reduced these individual measurements into high-precision multi-epoch relative astrometry, with the absolute calibration provided by 35 low-proper-motion 2MASS stars \citep{2003tmc..book.....C}. We derived the relative parallax and proper motion for \obj{b} using our standard MCMC approach and then converted to an absolute reference frame using the Besan\c{c}on galxy model to simulate the distances of the reference stars \citep{2003A&A...409..523R}.
%
Our eight epochs of astrometry spanning 3.16~years yield a parallax of $45.0\pm2.4$\,mas and proper motion of $(-286.1\pm1.3,-189.3\pm1.6)$\,\masyr, with reduced $\chi^2 = 1.15$ with 11~degrees of freedom (dof).

The central binary \obj{AB} was observed by the Pan-STARRS\,1 (PS1) telescope from 2009--2013, largely as part of the Pan-STARRS $3\pi$ Steradian Survey \citep{chambers2017}, with 9, 10, 16, 12, and 15 epochs in the \grizy\ filters, respectively. The $3\pi$ Survey was well-suited to parallaxes as every survey region was observed at opposition as well as evening and morning twilight. Astrometric calibration and automated calculation of parallaxes and proper motions are described by \cite{magnier2017.calibration}. 
The resulting astrometry is tied to the Gaia DR1 inertial system, with a correction for the proper motion bias introduced by Galactic rotation and solar motion. 
We re-analyzed \obj{AB}, adding one 2MASS and six PS1 epochs excluded from the automated analysis. 
The resulting astrometric solution is consistent with the automatic analysis, with slightly smaller errorbars.  
We used 53 likely quasars from \cite{2016ApJ...817...73H} within 1\degree\ of \obj\ to conclude that absolute corrections to the PS1 parallax and proper motions are negligible (mean quasar parallax of $0.3\pm3.5$\,mas and proper motions of $(0.1\pm1.4, 2.2\pm0.6)$\,\masyr).
Our final PS1 results give a parallax of $51.6\pm3.0$\,mas and proper motion of $(-272.0\pm1.7,-194.9\pm2.1)$\,\masyr, with reduced $\chi^2 = 2.0$ (125~dof).  We conducted the same analysis without the \gps\ data, which in principle could be most affected by photometric variability of the binary components, and found consistent results and uncertainties.

Our parallaxes from CFHT and PS1 are consistent, and we adopt the higher-precision CFHT parallax.  The two proper motions are inconsistent, especially in Right Ascension, which is plausibly due to \obj{b}'s orbital motion being incorporated into the CFHT data. We therefore adopt the PS1 proper motion, which also has the benefit of its reference frame being defined by quasars. 

\begin{figure}
\begin{center}
\includegraphics[scale=1.0,angle=0]{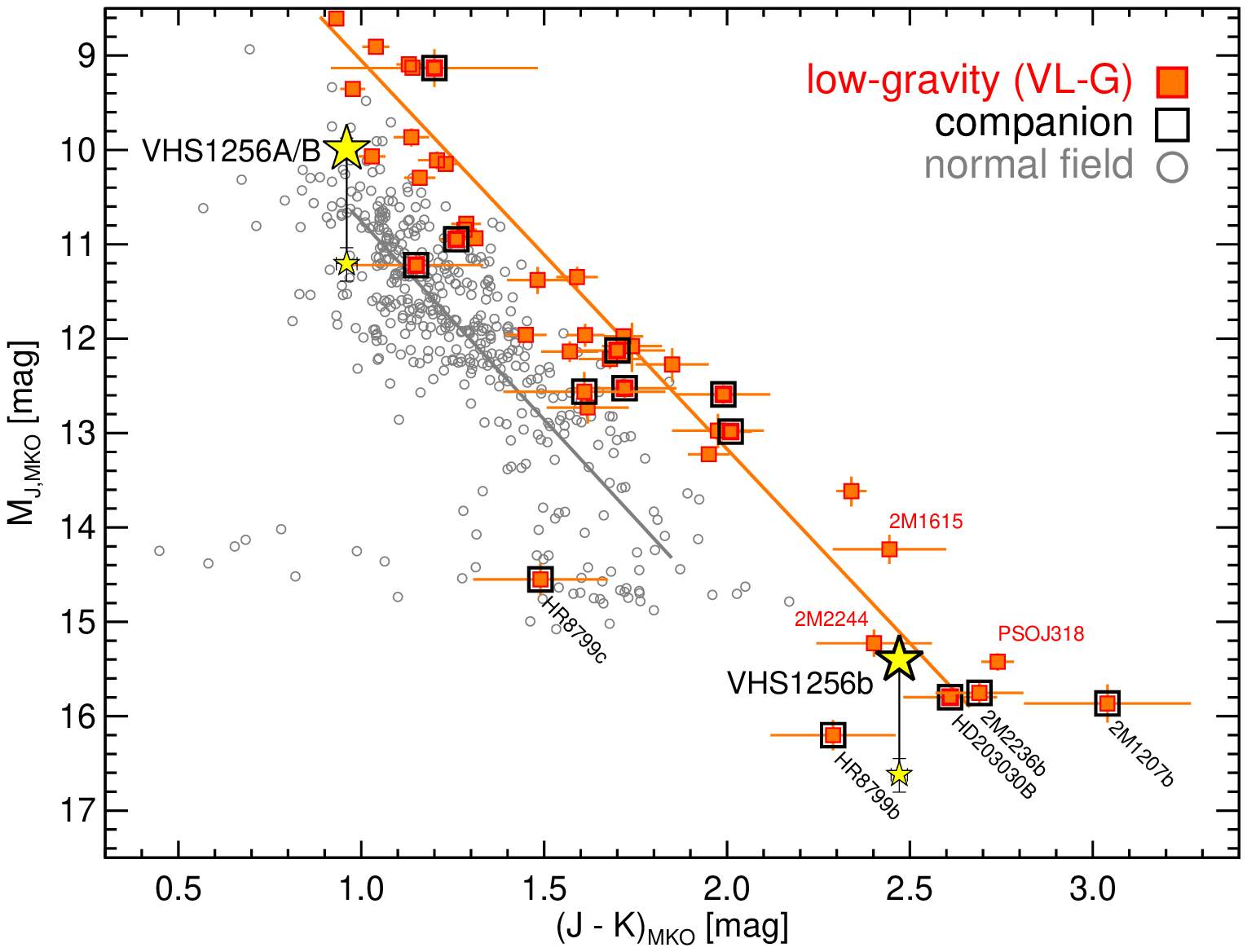}
\caption{Color--magnitude diagram (CMD) showing the old (small yellow stars) and new locations (large yellow stars) of the \obj\ components. (Deblended photometry is plotted for the equal-flux AB pair.)   The system is now more consistent with the locus of low-gravity objects.  Data for field and young objects come largely from \Gaia~DR2, Best \etal\ (in press), and \citet{2016ApJ...833...96L}.  (Also note that the first directly imaged planet 2MASS~J1207$-$39b \citep{2004A&A...425L..29C} appears brighter here than in previous plots in the literature because of its new DR2 parallax.) \label{fig:cmd}}
\end{center}
\end{figure}

Our parallaxes make all three components consistent with  known low-gravity (young) objects (Figure 1). Likewise, the infrared absolute magnitudes are in accord with young objects of the same spectral types from \citet{2016ApJ...833...96L}.

The system's new $UVW$ space motion and $XYZ$ location, from our astrometry and the \citet{2015ApJ...804...96G} radial velocity, continues to indicate that the system is not a member of known young moving groups, based on comparison with the \citet{2008hsf2.book..757T} groups and analysis with the \href{http://www.exoplanetes.umontreal.ca/banyan}{BANYAN $\Sigma$ webpage} \citep{2018ApJ...856...23G}.

We calculated component luminosities of $\log(\Lbol/\Lsun) = -2.94\pm0.07$, $-2.95\pm0.07$, and $-4.54\pm0.07$\,dex using photometry and spectral types from \citet{2015ApJ...804...96G}, our CFHT parallax, and the BC$_{K_S}$--SpT relation for young objects \citep{2015ApJ...810..158F}. For \obj{AB}, we derived properties from \citet{2015A&A...577A..42B} evolutionary models via rejection sampling following \citet{2018AJ....156...57D}, assuming a linear-uniform prior in age, a log-uniform prior in mass, and a conservative lithium depletion limit of $>$99.9\%, consistent with the  \ion{Li}{1} non-detection from \citet{2015ApJ...804...96G}. Assuming a maximum age of 300\,Myr, as in previous work, the resulting masses were each $94^{+10}_{-11}$\,\Mjup. The resulting age posterior had a 2$\sigma$ lower limit of 150\,Myr\ but otherwise mirrored our input prior. We used this output posterior as the input prior to estimate the properties of \obj{b} from the \citet{2008ApJ...689.1327S} hybrid evolutionary models, finding a mass of $19\pm5$\,\Mjup, $\Teff = 1240\pm50$\,K, and $\logg = 4.55^{+0.15}_{-0.11}$\,dex, all substantially higher than previous estimates. (For comparison, using the parallax from \citet{2015ApJ...804...96G} with our approach yields \obj{AB} masses of $70^{+2}_{-3}$\,\Mjup, a much narrower age posterior of 255--300\,Myr (2$\sigma$), and companion properties of $14.9^{+1.7}_{-1.4}$\,\Mjup, $\Teff = 960\pm30$\,K, and $\logg = 4.47^{+0.07}_{-0.05}$\,dex.)

In summary, our new parallax places \obj\ at $22.2^{+1.1}_{-1.2}$\,pc, raising the mass, temperature, and surface gravity estimates for the wide companion and moving the CMD position of the system's components into better agreement with previously known objects.

\begin{figure}
\centerline{\includegraphics[width=3.0in,angle=0]{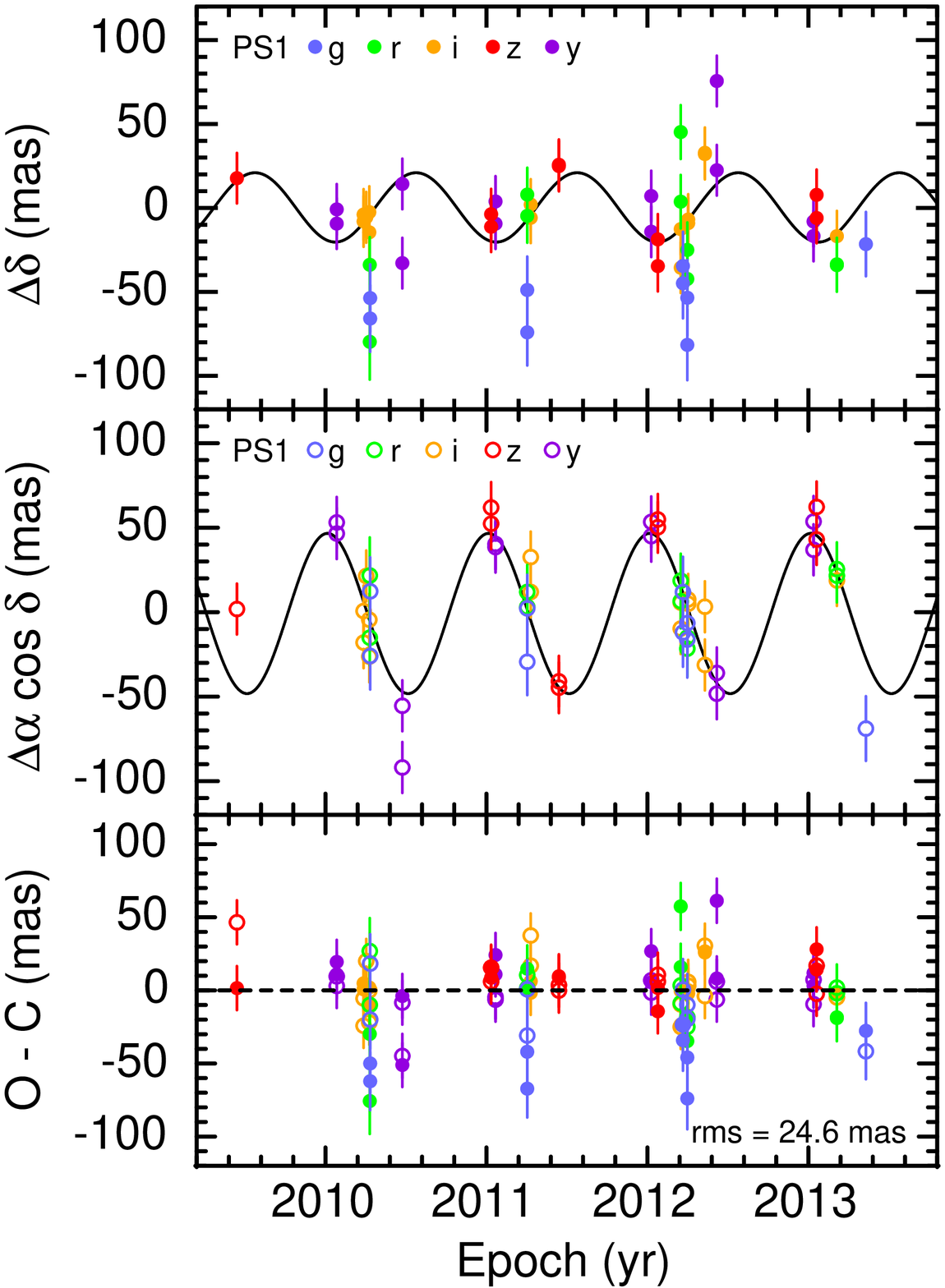}
\hskip 0.5in 
\includegraphics[width=3.0in,angle=0]{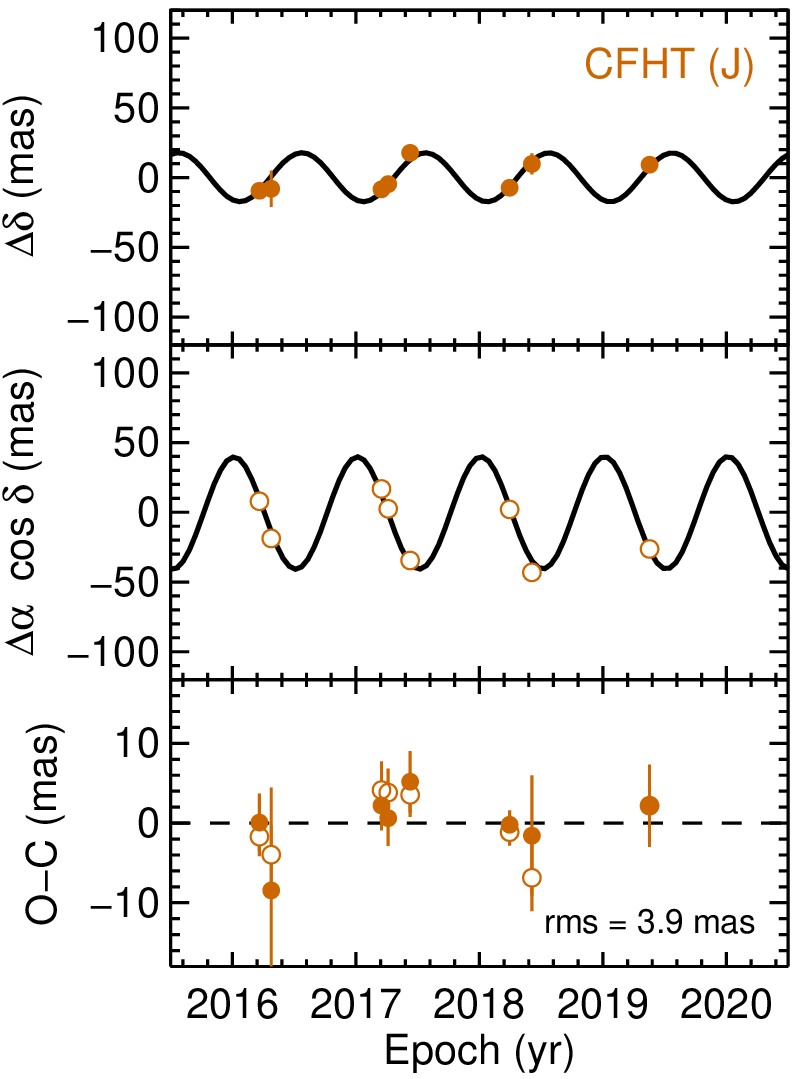}}
\caption{Relative astrometry of the inner binary \obj{AB} in integrated light from PS1 (left) and of the wide companion \obj{b} from CFHT/WIRCam (right). In the top two panels of each plot, the best-fit proper motions have been subtracted, and the best-fit parallax solution is plotted as a solid black line.  The bottom panel of each plot shows the residuals about the best-fit solution. \label{fig:plx}}
\end{figure}


\facilities{CFHT (WIRCAM), PS1 (GPC)}
\bibliographystyle{aasjournal}
\bibliography{refs.bib}

\end{document}